\documentstyle[12pt]{article}

\def\be{\begin{equation}}       
\def\ee{\end{equation}}         

\def\bea{\begin{eqnarray}}      
\def\eea{\end{eqnarray}}
                                
\def\ba{\begin{array}}
\def\ea{\end{array}}
\def\bd{\begin{displaymath}}
\def\ed{\end{displaymath}}

\def\eq{\begin{equation}}
\def\eqe{\end{equation}}
\def\eqa{\begin{eqnarray}}
\def\eqae{\end{eqnarray}}
\def\ena{\end{eqnarray}}

\def\eg{{\it e.g.~}}
\def\ie{{\it i.e.~}}

\def\Tr{{\rm Tr}}

\def\ra{\rangle}
\def\la{\langle}
\def\ud{{\mathrm{d}}}
\def\unit{1 \hskip-.3em \raise2pt\hbox{$ \scriptstyle |$ } }

\def\a{\alpha}
\def\b{\beta}
 
\def\d{\delta}
\def\e{\epsilon}           
\def\f{\phi}               
   
\def\g{\gamma}

\def\j{\psi}
\def\l{\lambda}
\def\m{\mu}

\def\o{\omega}  
\def\p{\pi}                
\def\r{\rho}                                     
\def\s{\sigma}                                   

\def\D{\Delta}

\def\G{\Gamma}

\def\cn{{\cal N}}
\def\co{{\cal O}}
\def\cp{{\cal P}}

\def\bop#1{\setbox0=\hbox{$#1M$}\mkern1.5mu
        \vbox{\hrule height0pt depth.04\ht0
        \hbox{\vrule width.04\ht0 height.9\ht0 \kern.9\ht0
        \vrule width.04\ht0}\hrule height.04\ht0}\mkern1.5mu}

\def\>{\rangle} 

\def\<{\langle} 
\def\Dsl{D \hskip-.6em \raise1pt\hbox{$ / $ } }
\def\to{\rightarrow}

\def\+{\oplus}

\def\Tr{{\rm Tr}\, }

\begin{document}
\begin{flushright}
MIT-CTP-3367 \\
{\tt hep-th/0305016}
\end{flushright}

\begin{center}
{\Large\bf
Instability and Degeneracy in the BMN Correspondence}
\vskip .6truecm

{\large\bf Daniel Z. Freedman${}^{\star\dagger}${}\footnote{
{\tt dzf@math.mit.edu}}\\ 
 and Umut G\"ursoy${}^{\dagger}${}\footnote{
{\tt umut@mit.edu}}} \\
\vskip 0.3truecm
 { *\it Department of Mathematics\\ 
*${}^{\dagger}$  Center for Theoretical Physics\\
 Massachusetts Institute of Technology\\
Cambridge MA 02139, USA }
\vskip 0.3truemm
\end{center}
\vskip .3truecm
\begin{abstract}
Non-degenerate perturbation theory, which was used to calculate
the scale dimension of operators on the gauge theory side of the
correspondence, breaks down when effects of
triple trace operators are included. We interpret this as 
an instability of excited single-string states in the dual string
theory for decay into the
continuum of degenerate 3-string states. We apply time-dependent
perturbation theory to calculate the decay widths from gauge
theory. These widths are new gauge theory data which can be
compared with future calculations in light cone string field theory.
\end{abstract}

\newpage

\section{Introduction}

The conjectured BMN correspondence \cite{bmn} potentially gives a 
perturbative, quantitative match between gauge theory and 
superstring theory. It was derived from the AdS/CFT
correspondence, but it probes string dynamics and 
thus circumvents the 
practical limitations of AdS/CFT to the supergravity
approximation to string theory. In the BMN correspondence,
one considers operators with large  $U(1)_R$ charge $J$. The
correlation functions of these operators are governed by two
parameters, the effective coupling
$\l' = g^2_{YM}N/J^2$ and genus expansion parameter $g_2 =J^2/N$,
and one works in the limit $N \to \infty,~~J \to \infty$ with
$J^2/N$ fixed. 

One family of operators commonly considered are
single- and multi-trace operators with two scalar ``impurity''
fields $\f (x), \j (x)$ and $J$ ``substrate'' fields $Z(x)$,

\be \label{op}
\co_n^{J_0}=\frac{1}{\sqrt{J_0N^{J_0+2}}}\sum_{l=0}^{J_0}e^{\frac{2\pi inl}{J_0}}
\Tr(\phi Z^l\psi Z^{J_0-l}).
\ee
and
\be\label{ops}
\co_n^{J_0,J_1,\ldots J_k}=:O_n^{J_0}O^{J_1}\ldots O^{J_k}:\;,
\ee
with $J = J_0+J_2+\cdots J_k.$ Here $O^{J}$ is the chiral primary
operator,
\be
\co^J=\frac{1}{\sqrt{JN^J}}\Tr(Z^J).
\ee

The states of superconformal $\cn =4$ gauge theory on
$R \times S^3$ which correspond to these operators\footnote {We use
the terms states and operators as if synomymous in the gauge
theory.} are dual to
states of the Type IIB superstring quantized in light-cone gauge
in the background pp-wave metric and 5-form:
\bea
ds^2 &=& -4dx^+dx^- - \mu^2 x^ix^idx^{+2}+dx^idx^i\;;
\qquad i=1,\dots, 8,\nonumber\\
F_{+1234} &=& F_{+5678} = \mu\times {\rm const}.\nonumber
\eea

The dual string states are those obtained from the ground state
$|p^+ \ra$ with light-cone momentum $p^+$ by the action of
creation operators $a^*_\f (n),~a^*_\j (n)$ with world-sheet momentum
$n$. More specifically one considers single- and multi-string
states:
\be \label{st}
a^*_\f (n) a^*_\j (-n)|p^+\>
\ee
and
\be \label{sts}
a^*_\f (n) a^*_\j (-n)|p_0^+\>\otimes
|p_1^+\>\otimes\cdots|p_k^+\>
\ee 
where light-cone momenta $p_i^+$ of individual string states are related 
to corresponding $R$-charges by $\m p_i^+\a'=\frac{J_i}{\sqrt{g^2_{YM}N}}$.    
The relation between the parameters of gauge theory and string
theory is:
\bea
\l ' &=& \frac{1}{(\m p^+ \a')^2} \\
g_2 &=& (4\p g_s)(\m p^+ \a')^2\\
g^2_{YM} &=& 4\p g_s
\eea

Despite the correspondence of notation the string states listed in
(\ref{st},\ref{sts}) are not the direct maps of the individual operators
in (\ref{op},\ref{ops}). The reason is that the operators mix through
nonplanar graphs \cite{bianchi,pots2,mit2} even in the free field theory whereas the
eigenstates of the free string Hamiltonian in (\ref{sts})
containing different numbers of strings are orthogonal. An
operator $S$ effecting a change of basis in the gauge theory has been
identified \cite{ucsb,vv} which makes the gauge theory states in (\ref{ops})
orthogonal in the free theory and it is the states obtained by
applying $S^{-1}$ to those of (\ref{ops}) which map
into the string states of (\ref{sts}). 

Quantitative tests of the correspondence are based on the
assumption that the light-cone Hamiltonian $P^- =H$ of string theory
corresponds to the field theory operator $\D -J$, the difference
between dilation and $R$-charge, through the relation
\be \label{dil}
\D - J = \frac{1}{\m} H
\ee
In planar order the eigenstates of $\D-J$ are the individual
states listed in (\ref{ops}), and the order $g_2^0$ eigenvalues
are 
\be \label{eig}
\o (n,J_0,J_1,..J_k) = 2 + \l ' \frac{n^2}{s_0^2}
\ee
with $s_0= J_0/J$. Operator mixing appears in order $g_2$, and 
to this order the $k$-trace eigenoperator acquires order $g_2$
admixtures of $(k \pm 1)$-trace operators \cite{pots2,mit2}.
The eigenvalue is first corrected in order $g_2^2$ to
\be \label{eig1}
\o (n,J_0,J_1,..J_k) \to 2 + \l ' [\frac{n^2}{s_0^2} + {g_2^2 s_0^2 
\over 4 \pi^2}( {1 \over 12} + {35 \over 32 \pi^2 n^2})  ] 
\ee
For $k=1$ the correction was found in \cite{pots2,mit2} 
and in \cite{umut2} for $k >1.$ For $k=1$, the correction exactly
matches the genus 1 energy shift of single-string states calculated
in light-cone string field theory \cite{psvvv,rsv}. This match
provides a basis independent test of the relation (\ref{dil}).
It has also been shown that related matrix elements of $H$ agree with
those of $\D - J$ after the basis change is made \cite{psvvv,caltech}.

The computations of (\ref{eig1}) in gauge theory used a formalism
equivalent to non-degenerate quantum mechanical perturbation
theory. Yet it is obvious from (\ref{eig}) that the zero order
eigenvalues of single-trace operators of momentum $n$ are
degenerate with multi-traces of momentum $m$ if $ns_0= \pm m$.
The $n=1$ state is non-degenerate, since $s_0=1$ would be required,
and $J_1,J_2,\cdots,J_k$ would vanish. But the $n=2$ single-trace
operator is degenerate with multi-traces with $m=\pm 2,~s_0= 1/2$. For
$n=3$ we have degeneracy with multi-traces when $m= \pm 2, ~s_0 =2/3$ and
when $m=\pm 1,~ s_0=1/3$, and so forth. 
One must thus question the validity of
non-degenerate perturbation theory, and this was discussed in
\cite{mit2}. One signal for breakdown of perturbation theory is
a divergence due to a vanishing energy denominator in the summation
formulas for shifts of eigenvalues.
It was pointed out in \cite{mit2} that perturbation theory would 
remain valid if the matrix elements in the numerator happened to
vanish at degeneracy, and that the matrix element of the
effective interaction between single- and double-trace states
does indeed so vanish. This is a necessary condition
for the validity of the order $g_2^2$ calculation leading to
(\ref{eig1}), but it is not sufficient. The single/triple mixing
matrix element is of order $g_2^2$. If it does not
vanish at degeneracy, it would also require \cite{mit2} the use
of degenerate perturbation theory with possible modification of
the result (\ref{eig1}) even though a divergence would not appear
until order $g_2^4$. 

An effective quantum mechanical formulation for the gauge theory,
which simplifies previous computations was developed in
\cite{golm1}. The single/triple matrix element was computed in this
formalism, and it {\bf does not} vanish at degeneracy. It is
this fact that motivates the present note in which we discuss 
the consequences for the physics of the BMN correspondence. In
string theory single-string states $|(n,-n)p^+\ra$ with $n>1$ would be
expected to be unstable, with decay to the continuum of
$(k+1)$-string states $|(m,-m)p_0^+,p_1^+,\cdots p_k^+\>$ in which
the total light cone momentum is divided continuously among the
$k+1$ strings. The lowest case $k=1$ should 
correspond to single/double trace mixing in gauge theory, and it has
been shown \cite{unp} that the relevant string theory
matrix element vanishes at order $g_2$ in agreement with the vanishing gauge
theory result mentioned above. Instability would then be expected
in string theory via a composite (order $g_2^2$) process in which the single
string first splits into two ``virtual'' strings and by a further
interaction into three final state strings. The non-vanishing
single/triple trace mixing matrix element, which is also
composite in a sense described below, is the signal of this
instability in gauge theory. 

More generally the gauge
theory has discrete single-trace states $\co^J_n$ embedded in a
continuum of multi-traces, since the ratios $J_i/J$ become
continuous variables with
range 0 to 1 in the BMN limit. It is a well known
phenomenon in quantum mechanics that a state which is purely in
the discrete sector at time $t=0$ undergoes irreversible decay to
the continuum of states which are degenerate in energy. This 
phenomenon can be derived using time-dependent perturbation
theory \cite{cotan}. The standard derivation must be
generalized in the present case because the relevant process is
composite.  We make this generalization and compute the order
$g_2^4$ contributions to the energy shift and decay width of
single trace states with $n \ge 2.$ The energy shift is the
less significant datum when there is instability, but it agrees
with the principal value calculation of \cite{golm1}. The value 
we find for the decay width would also emerge from the formalism of
\cite{golm1} if an $i\e$ prescription had been used rather than 
principal value (as suggested recently in \cite{sv}).
We show that the decay amplitude is invariant under the basis
change discussed above. It should therefore agree with the
amplitude computed in string theory.

The development of the continuous spectrum in the string theory
is intimately connected with the Penrose limit which produces the pp-wave
spacetime from ${\rm AdS}_5\times {\rm S}_5$. For finite radius
$R$ of ${\rm S}_5$, the null circle in $x^-$ is compact \cite{nas} and there
is a discrete spectrum of stable states. In the limit $R \to \infty$
the null circle becomes non-compact, producing a continuous
spectrum and the possibility of instability.

\newpage

\section{The Effective Gauge Theory Hamiltonian}

Let us denote an operator in the set (\ref{ops}) using the
generic notation $\co_i(x).$ With one-loop interactions included,
the general form of two-point functions is
\be \label{2pt}
\la \co_i(x) \co_j(y) \ra = {1 \over (x-y)^{2J+4}} [g_{ij} -
  h_{ij} \ln (x-y)^2M^2 ]
\ee
where $g_{ij}=g_{ji}$ is the free-field amplitude which defines a
positive-definite inner product or metric, and $h_{ij}=h_{ji}$
describes the order $\l' = g_{YM}^2N/J$ interactions. We use a
real notation for simplicity, but it is accurate since the
correlators of our operators are real. The diagonal elements of
$g_{ij},~h_{ij}$ are of order $1 + \co(g_2^2)$ while off-diagonal
elements between operators containing $k$ and $k'$ traces are of
order $g_2^{|k-k'|}$.

One diagonalizes this system via the relative eigenvalue problem
\be \label{rel}
h_{ij} v^j_\a = \g_\a g_{ij} v^j_\a
\ee
There is a complete set of eigenvectors $v^i_\a$ which are
orthogonal with respect to the metric, viz. $v^i_\a g_{ij} v^j_\b
= \d_{\a\b}$. The system (\ref{rel}) is equivalent to the
conventional eigenvalue problem for the ``up-down Hamiltonian''
$h^i_j = g^{ik} h_{kj}$, namely
\be \label{con}
h^i_j v^j_\a = \g_\a v^i_\a,
\ee
but one must remember that $h^i_j$ is not naively Hermitean
(symmetric), but Hermitean with respect to $g_{ij}$,
i.e. $g_{ik}h^k_j= h^k_ig_{kj}$.

It is easy to show that the eigenoperators are $\co_\a (x) =
v^i_\a \co_i(x)$ and have diagonal 2-point functions:
\bea \label{eigop}
\la \co_\a (x) \co_\b(y) \ra &=& {\d_{\a\b} \over (x-y)^{2J+4}}   
[1 - \g_\a \ln (x-y)^2M^2]\\
&\sim & { \d_{\a\b} \over (x-y)^{2\D_\a}},
\eea
from which one can read the scale dimension $\D_\a = J+2+\g_\a$.

A simpler method for field theory computations was developed in
\cite{golm1}. It is a method to compute $h^i_j$ directly with no
need for the complicated combinatorics of Feynman diagrams
required in earlier work. The resulting effective Hamiltonian has
a quite simple and striking structure. The method applies to
2-point functions of the operators of (\ref{ops}) and has
recently been extended \cite{golm2} to a wider class of scalar
operators. 

We refer readers to \cite{golm1} for an explanation of the method,
and we begin our discussion with (11) of that paper. Certain 
``end-point terms'' and other terms which vanish in the BMN limit
are neglected in (11), and we note that the omitted terms
actually vanish in the channel which is symmetric under exchange
$\phi \leftrightarrow \psi$ of the two impurities, so (11) is
exact in this channel. Although the BMN limit is taken at an
early stage in \cite{golm1}, we use a discrete finite $J$
version of the method and take the limit $J \to \infty$ at the
final stage of computation.

In this method $h^i_j$ is replaced by matrix elements of an
effective Hamiltonian $H= H_0 + H_+ + H_-$. The action of H on
gauge theory states/operators is given in (11) of \cite{golm1} in
which the operators contain impurities of fixed spacing $l$, \ie
$\co^{J_0,J_1\ldots J_k}_l \sim \rm{Tr}(\phi Z^l \psi Z^{J_0-l})
\rm{Tr}Z^{J_1}\ldots \rm{Tr} Z^{J_k}$. After a discrete Fourier
transform with respect to $l$, which is equivalent to that in
(\ref{ops}) for large J, one obtains the following equations:
\bea \label{heff}
H_0\co_m^{J_0,J_1,\cdots,J_k}&=&\l '\frac{m^2}{s_0^2}
\co_m^{J_0,J_1,\cdots,J_k}\nonumber\\
H_+\co_m^{J_0,J_1,\cdots,J_k}&=&\l
'g_2\sum_{\a'_{k+1}}V_{\a_k}^{\a_{k+1}}\co_{m'}^{J_0-J_{k+1},J_1,
\cdots J_{k+1}}\\
H_-\co_m^{J_0,J_1,\cdots,J_k}&=&\l
'g_2\sum_{i,\a'_{k-1}}V_{\a_k}^{\a_{k-1}}\co_{m'}^{J_0+J_i,J_1,\cdots 
J_i\!\!\!\!\backslash,\cdots J_k}\nonumber
\eea
where we introduce a collective index notation in the matrix
elements $V_{\a_k}^{a_{(k \pm 1)}}$, namely
$\a_k=\{m,s_0,\dots,s_k\}$. Here $s_i=J_i/J$ satisfy $\sum_i s_i =1$.  

The right sides of the equations for $H$ define
contributions to matrix elements $h^i_j$. Symbolically,
the structure is $H \co_j = h^i_j \co_i$. Note that the
interaction terms are purely of order $g_2$ and
describe the splitting/joining of a $(k+1)$-trace operator into
superpositions of $(k+1\pm 1)$-trace operators. The Hamiltonian
$\tilde{H} = S H S^{-1}$ transformed to string basis agrees
\cite{sv} with the Hamiltonian of string bit formalism
which contains order $g_2$ splitting/joining interaction and an
order $g_2^2$ contact term. 

For large finite $J$ the matrix elements are
\bea
V_{\a_k}^{\a'_{k+1}}&=&
\frac{1}{\pi^2\sqrt{J}(k+1)!}\sqrt{\frac{s_0-s'_0}{s_0s'_0}}(\frac{m'}{s'_0})
\frac{\sin^2(\pi
  ms'_0/s_0)}{\frac{m'}{s'_0}-\frac{m}{s_0}}\Delta_{ss'}\label{Vcoef}\\
V_{\a_{k+1}}^{\a'_{k}}&=&\frac{1}{\pi^2\sqrt{J}k!}
\sqrt{\frac{s'_0-s_0}{s_0s'_0}}(\frac{m'}{s'_0})\frac{\sin^2(\pi
  m's_0/s'_0)}{\frac{m'}{s'_0}-\frac{m}{s_0}}\Delta_{s's}\label{Vcoef2} 
\eea
where $\Delta_{ss'}$ is a product of delta functions,
\be\label{delta}
\Delta_{ss'}= \sum_{P\in S_{k+1}}\d_{s_1,s'_{P(1)}}
\cdots\d_{s_{k},s'_{P(k)}}\d_{s_0,s'_0+s'_{P(k+1)}}.
\ee

It is in this form that we will use these equations. In the BMN
limit, matrix elements we calculate agree with those of the
continuum formulation of (\cite{golm1}).

We have also obtained a version of (\ref{heff}) valid at any
finite $J$ in the $\phi \leftrightarrow \psi$ symmetric channel.
Eigenoperators of $H_0$ are superpositions of those of fixed
spacing, namely
\be
\co^{J_0,J_1\ldots J_k}_n = \frac{1}{\sqrt{J_0+1}}\sum_{l=0}^{J_0}
\cos({\pi n (2l+1) \over J_0+1}) \co^{J_0,J_1\ldots J_k}_l
\ee
These operators have eigenvalues of $H_0$ given by ${g_{YM}^2 N
\over \pi^2} \sin^2({\pi n \over J_0 +1}) $ which approaches the
eigenvalue in (\ref{heff}) as $J_0 \to \infty$. These results agree
with those of \cite{beisert, minahan}. The interaction terms
still describe order $g_2$ splitting/joining of traces, but they are
more complicated than those of (\ref{heff}). For example when
$H_+$ acts on the the single-trace $\co^J_n$ one obtains a
superposition of double-trace operators $\co^{J_0,J_1}_m$ where
$J_0=J-J_1$ with expansion coefficients
\be
{g^2_{YM}\sqrt{J_1} \over \pi^2 \sqrt{(J+1)(J_0+1)}} \sin({\pi m \over
J_0+1}) A(n,m,J_1)
\ee
\be
 A(n,m,J_1) = \frac{1}{2} \sin({ \pi n J_1 \over J+1})\{[\frac{ \sin( \pi
({ n J' \over J+1} - {m \over J_0+1}))} {\sin (\pi({n \over J+1} +
{m \over J_0+1}))}] - [ m \to -m] \}
\ee
One notes that these coefficients exactly reduce to (\ref{Vcoef}) in
the symmetric channel in the $J\to\infty$ limit. 
Another thing to note here is that working at finite $J$ does not
resolve the degeneracy problem. For example, single- and
double-trace operators are degenerate when $\frac {n}{J+1} = \pm
\frac {m}{J_0+1}$, and there is a similar degeneracy condition
for $(k+1)$-trace operators. One may also observe that
$ A(n,m,J')$ does not vanish at degeneracy for finite $J$,
although it does vanish as $J \to \infty$. In principle, one
should apply degenerate perturbation theory to the calculation of
eigenvalues of $H$ even before triple-trace operators are
included. However, we will ignore this complication, which
very likely disappears in the large $J$ limit.

Our primary goal is to apply time-dependent perturbation theory
to study the time evolution of a state which is purely
single-trace at time $t=0$. We will calculate the decay rate of
such a state into degenerate triple-trace states. 
We have
found it useful to illustrate the essential physics in quantum
mechanical toy models and then adapt the results in the models
to the BMN limit of the Hamiltonian (\ref{heff}).

\section{Quantum Mechanical Models} 

Our calculation of the decay rate is based on the treatment
of the decay of a discrete state embedded in a continuum in
\cite{cotan}. This treatment needs to be modified for our case,
but we first review it to set the basic technique in the
context of the present up-down matrix formalism.

We thus consider a quantum mechanical system with a set of
discrete states $|n \>$ and continuum states $|\a \ra$ where
$\a$ denote the continuous labels of the state.
The Hamiltonian is $H= H_0 + V$ where $H_0$ and $V$ are given
by the following matrices: 
\bea
H_0&=&\left(\begin{array}{cc}
E_i\d^i_j & 0 \\
0 & E_\a\d(\a-\b)  
\end{array}\right),\\
V&=&\left(\begin{array}{cc}
0 & V^i_\a \\
V^\a_i & 0
\end{array}\right).
\eea
Note that $V$ has vanishing matrix elements between pairs of
discrete or pairs of continuum states in agreement with the
interactions $H_{\pm}$ of (\ref{heff}). 
We look for the solution of the Schr\"odinger evolution equation
$ i \frac{d}{dt} |\Psi(t) \ra  = H |\Psi(t) \ra$, with $|\Psi(t) \ra$
given by the formal vector: 
\bd
|\Psi(t) \ra=\left(\begin{array}{c}
a^i(t)e^{-i\,E_i\,t} \\
\f^a(t)e^{-i\,E_a\,t}
\end{array}\right).
\ed
and the initial conditions $a^m(0) = \d^m_n,~~ \phi^\a(0) =0$.

It is easy to see that the discrete and continuum components of
$|\Psi(t) \ra$ satisfy:
\bea
i {d \over dt} a^n(t) &=& \int d\a V^i_\a \exp i(E_n -E_\a)t~ 
\phi^\a(t)\\
i {d \over dt} \phi^\a(t) &=& \sum_{m} V^\a_m \exp i(E_\a-E_m)t~
a^m(t)
\eea
We integrate the second equation using the initial conditions
above and substitute the resulting expression for $ \phi^\a(t)$
in the first equation, obtaining an equation involving only the
discrete components, namely
\be \label{uncoup}
{d \over dt} a^n(t) = -\int d\a \sum_{m} \int_0^t dt' 
\exp i(E_n-E_\a)t~\exp
i(E_\a -E_m)t'~V^n_\a V^\a_m~a^m(t')
\ee
Next we separate the energy variable in the continuum measure by
writing $d\a =dE d\b \r(\b,E)$ and define the matrix
\be \label{kdef}
K^n_m(E) = \int d\b  \r(\b,E)~V^n_\a V^\a_m.
\ee
We assume that $K^n_m(E)$ is a slowly varying function of
energy whose scale of variation is $\D E$. Substituting
(\ref{kdef}) in (\ref{uncoup}), we find
\be \label {unk}
{d \over dt} a^n(t) = -\int_{0}^{\infty} dE K^n_m(E)
~\exp i(E_n-E_\a)t~\int_{0}^{t} dt'~\exp i(E_\a -E_m)t' a^m(t'). 
\ee

We now follow \cite{cotan} and make the short-time approximation
$a^n(t') \approx 1,~~ a^m(t') \approx 0$ for $m \ne n$ in
(\ref{unk}). We encounter the well known integral
\bea \label{wno}
\int_{0}^{t} dt' \exp i(E_n-E)(t-t') &=& \frac{\exp i(E_n-E)t
~-1}{i(E_n-E)}\nonumber\\
{}&\approx& \pi \d(E_n-E) + i\cp({ 1 \over E_n-E}).
\eea
The last result is strictly correct in the limit $t \to \infty$,
but it is effectively valid within integrals of functions f(E)
for times much larger than the inverse scale of variation $\D E$.

This approximate treatment of (\ref{unk}) thus leads to the
result 
\bea \label{toy}
a^n(t) &\approx & 1 - (\frac{\G_n}{2} + i\D \o_n)t \\
   \G_n &=& 2 \pi K^n_n(E=E_n) \\
  \D \o_n &=& \cp \int_{0}^{\infty} dE {K^n_n(E) \over  E_n-E}.
\eea 
The quantities $ \G_n, ~\D \o_n$ are interpreted as the decay
width and energy shift of the unstable state to lowest order in
the interaction $V$. The approximations made in the derivation
are valid under the conditions: $1/\D E << t << 1/\G_n$, \ie the
time t must be long compared to the inverse scale of variation
of $K^n_n(E)$ and sufficiently short to justify the short-time 
approximation to (\ref{unk}). Note that it was not necessary to
specify a scalar product in Hilbert space. It is necessary to
define the measure $d\a =dE d\b \r(\b,E)$ explicitly. In our
problem this measure is determined by the BMN limit of the
discrete formulation. 

There are additional checks of the self-consistency of the
method.  One can show that the components $a^m(t),~~m \ne n$
satisfy  $a^m(t) \sim t^2$ for small $t$, and that the unitarity
constraint $|a^n|^2 + \int d\a |\phi^\a| = 1$ is satisfied to
order $t$.

Applied to our problem the
treatment above gives the result $\G_n=0$, since $K^n_n(E_n)$
vanishes at degeneracy\footnote{It is curious that $K^n_n(E) \le 0$
  because $V_n^{\a_2}$ is not Hermitean.}. 
This is because triple-trace states enter
the dynamics only at higher order in the coupling $g_2$. The toy
model above must be generalized to include this effect. Note that 
the energy shift $ \D \o_n$ in (\ref{toy}) agrees with the order
$g_2^2$ contribution to the scale dimension of single-trace BMN
operators in (\ref{eig1}).
 
The generalized model includes three types of states: i)~the
discrete ``single-trace'' $|n \>$, ii)~ continuous
``double-trace'' $|\a_2 \>$, and iii)~ continuous
``triple-trace'' $|\a_3 \>$. The free Hamiltonian $H_0$ is
diagonal with energy eigenvalues $E_n, E_{\a_2}$, and $
E_{\a_3}$, respectively. 
The interaction matrix and time-dependent state vector which
generalize those of the model above are:
\bea \label{int}
V&=&\left(\begin{array}{ccc}
0 & V^i_{\a_2} & 0 \\
V^{\a_2}_i & 0 & V^{\a_2}_{\a_3}\\
0 & V^{\a_3}_{\a_2} & 0
\end{array}\right),\\
|\Psi(t) \ra&=&\left(\begin{array}{c}
a^i(t)e^{-i\,E_i\,t} \\
\f(\a_2,t)e^{-i\,E_{a_2}\,t}\\
\s(\a_3,t)e^{-i\,E_{a_3}\,t}
\end{array}\right).
\eea
Again we need the solution of the Schr\"odinger equation $ i d / dt
|\Psi(t) \ra  = (H_0+V) |\Psi(t) \ra $ with initial conditions:
$a^m(0) = \d^m_n,~\phi^{\a_2}(0)=0, \s^{\a_3}(0)=0$. The equations
linking the components of $|\Psi(t) \ra$ are
\bea
i {d \over dt} a^n(t) &=& \int \!\!\ud\a_2 V^n_{\a_2} e^{iE_{n\a_2}t}~ 
\phi^{\a_2}(t)\label{one}\\
i \frac{d}{dt} \phi^{\a_2}(t) &=& \sum_m V^{\a_2}_m  e^{iE_{\a_2m}t}~
a^m(t) + \int \!\!\ud\a_3 V^{\a_2}_{\a_3}~e^{
iE_{\a_2\a_3}t}~ \s^{\a_3}(t)\label{two}\\
i \frac{d}{dt} \s^{\a_3}(t) &=& \int \!\!\ud\a_2  V^{\a_3}_{\a_2}~e^{
iE_{\a_3\a_2}t}~  \phi^{\a_2}(t) \label{three}
\eea
We use the notation $E_{n\a_2} = E_n - E_{\a_2}$, etc. for
differences of energy.

We now wish to process the information in (\ref{one}-\ref{three})
to obtain a relation describing the coupling of the discrete
components of $|\Psi(t) \>$ alone. Rather than the exact equation
(\ref{uncoup}) in the simpler model, we will obtain a relation
which is accurate to fourth order in the potentials. For this
purpose we begin in 
straightforward fashion to integrate (\ref{three})
obtaining an expression for $\s^{\a_3}$ in terms of
$\phi^{\a_2}$. We then substitute this in (\ref{two}) and
integrate that, and substitute the result in the first equation
which becomes
\bea \label{mess}
 {d \over dt} a^n(t) &=&  -\int\!\!\ud\a \sum_{m'} \int_0^t dt' 
e^{iE_{n\a_2}t+iE_{\a_2m}t'}~V^n_\a V^\a_{m'}~a^{m'}(t')\\ 
{}&+&\!\!\!\!\!i\!\!\int\!\!\ud\a_2\,\ud\a_3\,\ud\a'_2 \int_{0}^t\!\!\ud t'\!\!\int_{0}^{t'}
\!\!\ud t'' e^{iE_{n\a_2}t+iE_{\a_2\a_3}t'+iE_{\a_3\a'_2}t''}
V^n_{\a_2}V^{\a_2}_{\a_3}V^{\a_3}_{\a'_2}\phi^{\a'_2}(t'')\nonumber
\eea

The next step is to substitute for $\phi^{\a'_2}(t'')$
in the equation above the value obtained by integrating the first term
of (\ref{two}) with ``source'' $a^m$. The term with
``source'' $\phi^{\a_2}$ is of higher order in the potentials
through (\ref{three}) and can be dropped. This gives us the net
contribution of triple-trace intermediate states, denoted by
\bea \label{trip}
{d \over dt} a^n(t)_{triple} &=& \sum_{m}
\int\!\!\ud\a_2\,\ud\a_3\,\ud\a'_2 \int_{0}^t\!\!\ud 
 t' \int_{0}^{t'} \!\!\ud t'' \int_{0}^{t''}\!\!\ud t'''
e^{iE_{n\a_2}t+iE_{\a_2\a_3}t'} \nonumber\\
{}&&\; e^{iE_{\a_3\a'_2}t''+iE_{\a_2m}t'''}~
V^n_{\a_2}V^{\a_2}_{\a_3}V^{\a_3}_{\a'_2}V^{\a'_2}_m~a^m(t''') 
\eea

To obtain the contribution of single-trace intermediate states 
at order $g_2^4$,
we integrate the first term of (\ref{mess}) to find an expression
for $a^n(t)$ (with $n \to m$). This expression is then reinserted
for $a^{m'}(t')$ in (\ref{mess}) to obtain the iterated contribution
\bea \label{sing}
{d \over dt} a^n(t)_{single} &=& \sum_{m,m'}
\int\!\!\ud\a_2\,\ud\a'_2 \int_{0}^t\!\!\ud
 t' \int_{0}^{t'} \!\!\ud t'' \int_{0}^{t''}\!\!
\ud t'''e^{iE_{n\a_2}t+iE_{\a_2m}t'}
\nonumber\\
{}&&\;e^{iE_{m\a'_2}t''+iE_{\a_2m'}t'''}~
V^n_{\a_2}V^{\a_2}_{m}V^{m}_{\a'_2}V^{\a'_2}_{m'}~a^{m'}(t''')
\eea
The full expression for $\frac{d}{dt}a^n(t)$ to fourth order is
\bea \label{full}
 {d \over dt} a^n(t) &=&  -\int\!\!\ud\a_2 \sum_{m'} \int_0^t dt' 
e^{iE_{n\a_2}t+iE_{\a_2m}t'}~V^n_{\a_2} V^{\a_2}_n \nonumber\\
{}&&+ ~\frac{d}{dt}a^n(t)_{triple}~ +~\frac{d}{dt}a^n(t)_{single}
\eea
The first term is just the short-time approximation to (\ref{unk}) in the simple model
in different notation. We will not discuss it further since it
does not contribute to the fourth order amplitudes of primary
concern in the generalized model.

We now make the short-time approximation $a^n(t') \approx 1,~~
a^m(t') \approx 0$ for $m \ne n$ in (\ref{trip}) and
(\ref{sing}). We will present the treatment of the triple-trace part
explicitly and then summarize the rather similar steps needed for
the single-trace part.

The nested set
of time integrals in (\ref{trip}) can easily be done, and the
result (including the overall factor  $\exp i(E_{n\a_2})t)$ is
\be\label{tint}\frac{e^{iE_{n\a_3}t}-1}{E_{\a_3n}E_{\a_2\a_3}}
-\frac{e^{iE_{n\a_2}t}-1}{E_{\a_2 n}E_{\a_2\a_3}}
+\frac{e^{iE_{n\a'_2}t}-e^{iE_{n\a_2}t}}{E_{\a_2\a'_2}E_{\a_3\a'_2}}
+\frac{e^{iE_{n\a_2}t}-e^{iE_{n\a_3}t}}{E_{\a_2\a_3}E_{\a_3\a'_2}}
\ee
In an obvious fashion we subtract and add $1$ in the last two
terms of (\ref{tint}). We thus obtain six terms with the
structure $\exp iEt~-1$ divided by energy denominators.  
In our discussion of the contribution of these six terms to
(\ref{trip}) in the short-time limit, we need the fact that all the
interaction matrix elements of the actual problem vanish at
degeneracy due to the trigonometric factors in (\ref{Vcoef}).  
We assume the same property in the toy model so that all energy
integrals which appear when (\ref{tint}) is
inserted in (\ref{trip}) converge despite the singular
denominators. 

We apply the $i\e$ prescription of the last line of (\ref{wno}) in
each of the six terms, multiplying and dividing by the energy
denominators which are missing in the last four terms. It is easy to
see that all contributions cancel pairwise in third and fourth and in the
fifth and sixth terms. In the first and second terms we obtain
\bea \label{trip1}
\frac{d}{dt}a^n(0)_{triple} &=& -\int\!\!\ud\a_2\,\ud\a_3\,\ud\a'_2
V^n_{\a_2}V^{\a_2}_{\a_3}\frac{1}{E_{\a_2\a_3}}
V^{\a_3}_{\a'_2}V^{\a'_2}_n\frac{1}{E_{\a'_2n}}\nonumber\\
{}&&\;[\pi \d (E_{n\a_3}) +i\cp (\frac{1}{E_{n\a_3}}) - \d(E_{n\a_2})
-i\cp(\frac{1}{E_{n\a_2}})]\nonumber\\
\eea
We now note that the term with  $\d(E_{n\a_2})$ vanishes because
$V^n_{\a_2}=0$ at degeneracy. Because of this vanishing one can
combine the two principal value terms without ambiguity. The
triple-trace contribution to the decay amplitude can then be
written as 
\be \label{tricon}
\frac{d}{dt}a^n(0)_{triple} = -\int d\a_3 U^n_{\a_3} 
[\pi \d (E_{\a_3 n}) ~+~i\cp (\frac{1}{E{\a_3 n}})]  U^{\a_3}_n
\ee
where we have introduced the effective composite interactions
which couple single- and triple-traces, namely
\bea 
U^n_{\a_3} &=&  \int d\a_2 \frac{V^n_{\a_2} V^{\a_2}_{\a_3}}{E_{n\a_2}}\label{mel1}\\
U^{\a_3}_n &=& \int d\a_2 \frac{V^{\a_3}_{\a_2} V^{\a_2}_n}{E_{n\a_2}}\label{mel2}.
\eea

The single-trace contribution (\ref{sing}) can be treated
similarly. One makes the short-time replacement $a^m(t''') \to
\d^m_n$. The time integrals are then easily done, but separately
for the two cases $m \ne n$ and $m = n$. The contributions of the
various terms to $\frac{d}{dt} a^n(t)$ at small $t$ are then
analyzed as above. In each case there is one term which contributes
at $t=0$ via the $i\e$ prescription. In each there is a
$\d(E_{n\a_2})$ which drops because $V^n_{\a_2}=0$. The
contributions of two (non-singular) principal
value integrals remain in the final result
\bea \label{sing1}
\frac{d}{dt} a^n(0)_{single} &=&i\sum_{m} U^n_m
\frac{1-\d_{E{nm}}}{E_{mn}} U^m_n \\
{}&& -i U^n_n \int d\a_2 \frac{V^n_{\a_2}V^{\a_2}_n}{E_{\a_n}^2}
\eea
which is largely written in terms of the composite interaction
\be \label{mek}
U^n_m =\int d\a_2 \frac{V^n_{\a_2} V^{\a_2}_m}{E_{n\a_2}}
\ee

We can now interpret the results (\ref{trip1},\ref{sing1}) in
terms of (\ref{toy}). We find the decay width
\be \label{wid}
\G_n = 2\pi\int d\a_3  \d (E_{\a_3 n}) U^n_{\a_3}U^{\a_3}_n
\ee
and energy shift
\bea \label{eshif}
\D\o_n &=& U^n_n - \sum_{m} U^n_m\frac{1-\d_{E_{nm}}}{E_{mn}}
U^m_n\nonumber\\
{}&& + U^n_n \int d\a_2 \frac{V^n_{\a_2}V^{\a_2}_n}{E_{\a_2 n}}\nonumber\\
{}&& -\int d\a_3 U^n_{\a_3}\cp (\frac{1}{E_{\a_3 n}}) U^{\a_3}_n
\eea

We are primarily interested in the decay width, which will be
evaluated for the BMN system (\ref{heff}, \ref{Vcoef}) in section 5.
However, we note that the energy shift agrees with the result of
fourth order non-degenerate (time-independent) perturbation
theory for any quantum-mechanical system whose state space and interaction 
structure are the same as the present model as defined in (\ref{int}) and the
discussion above it\footnote{  Most treatments of perturbation
theory assume a hermitean Hamiltonian, but the standard formulas remain
valid  when rewritten in terms of $h^i_j$}. Of course, the principal
value derived here to resolve the divergence in the last
term is not present in conventional perturbation theory. The
physical interpretation of our result is that the pole
of the resolvent operator $1/(H_0-E)$ at $E=E_n$ due to the discrete
state $|n \ra$ is shifted to $E = E_n + \D\o_n - i\G/2$ in the
complex plane by the interaction in $1/(H_0+V-E)$. The state  $|n
\ra$ is unstable in the full theory.

The goal of the Hamiltonian formulation of \cite{golm1} was the calculation
of anomalous dimensions of BMN operators. In our opinion the calculations
undertaken for this purpose should be revised to incorporate
degenerate rather than non-degenerate perturbation theory. One
may note that the various contributions to our energy shift
(\ref{eshif}) agree exactly with those of (25) of
\cite{golm1}. So the result there should be interpreted as the
real part of the shift of a pole rather than the anomalous
dimension of an eigenstate of the dilatation operator. 

It is interesting that the present time-dependent treatment
provides justification for the recent suggestion in \cite{sv} of an
$S$-matrix approach to the BMN system (see also \cite{sj}) 
which would require an
$i\e$ prescription in the genus two calculations of \cite{golm1}.
The idea of an $S$-matrix is fully compatible with our interpretation of
the instability of the states $|n \ra$.


\section{Basis Independence}

It is our intention to derive a formula for the decay widths
which can be compared with future calculations in light-cone
string field theory. We must therefore determine the effect of
the change to the basis in which gauge theory and string theory
calculations should match. It turns out that the result is not
changed by the basis transformation (at least to order $g_2^4)$.
Basis independence would be a triviality if it were implemented
as a standard ``change of representation'' in quantum mechanics,
since matrix elements are invariant. However, things are not
entirely trivial since states and operators actually transform
contragrediently. We follow the recent paper \cite{sv}, although the
same result should emerge from the similar formalism of
\cite{caltech}.

The metric $g_{ij}$ defined by the free two-point functions of the $\co_i$
of (\ref{2pt}) can be referred to local frames using the vielbein
\be
 g_{ij} = e^k_i \d_{kl} e^l_j
\ee
We found the usual practice of different fonts for the
``frame'' and ``coordinate'' indices confusing in the present
application, so we prefer to emphasize that the same physical
variables are indexed by both upper and lower indices. The
inverse vielbein is denoted by $f^i_k$, so that \eg $f^i_k e^k_j
=\d^i_j$. The operators with diagonal free two-point functions are
\be \label{oops}
{\tilde \co}_k = f^i_k \co_i \equiv S^{-1} \co_k
\ee
where we have defined the Hilbert space operator $  S^{-1}$ by
the last equality. Our  $  S^{-1}$ has exactly the properties of
$S^{-1/2}$ in \cite{sv}. In particular the string basis
Hamiltonian is 
\be \label{sham}
{\tilde H} = S H S^{-1}
\ee
which was shown to be exactly the Hamiltonian of the string bit
formalism \cite{v,vv} whose order $g_2$ splitting/joining
interaction agrees with string field theory and order $g_2^2$
contact term also agrees (if a certain truncation of intermediate
states is made) \cite{rsv}.

The toy models of Section 3 can now be described in a new
notation in which the states $\co_i$ and ${\tilde \co}_j$ span different 
bases of the Hilbert space, the gauge theory basis and the string
basis, respectively. The time evolution problems treated in
Section 3 as models for gauge theory can be described as
follows. We found (approximately)
the state
\be \label{state1}
|\Psi(t)\>= \sum_{i} a^i(t) \co_i
\ee
which evolves with time by the Hamiltonian $H$ and satisfies an
initial condition $a^i(0) = 1$ for $i=i_0$ and $a^i(0)=0$ for $i
\ne i_0$. From the Schr\"odinger equation $i \frac{d}{dt}|\Psi(t)\>
= H |\Psi(t)\>$ we easily derive the component
evolution
\be \label{comp1}
i  \frac{d}{dt}  a^i(t) = H^i_j a^j(t).
\ee

In string basis we would instead be interested in the
state
\be \label{state2}
|{\tilde \Psi}(t)\> = \sum_{i} {\tilde a}^i(t) {\tilde \co}_i
\ee
which evolves in time by the Hamiltonian ${\tilde H}$ with the
initial conditions  ${\tilde a}^i(0) = 1$ for $i=i_0$ and
${\tilde a}^i(0)=0$ for $i \ne i_0$.

We now attempt to relate the time-dependent expansion
coefficients  $a^i(t)$ and ${\tilde a}^i(t)$. The Schr\"odinger
equation  $i \frac{d}{dt}|{\tilde \Psi}(t)\> = {\tilde H} |{\tilde
\Psi}(t)\>$ can be expanded as   
\be \label{sch2}
i  \frac{d}{dt} \sum_{i} {\tilde a}^i(t) S^{-1} \co_i = \sum_{m} 
 {\tilde a}^i(t)  {\tilde H}S^{-1} \co_i.
\ee
We apply $S$ to both sides and obtain the component equations
\bea \label{comp2}
i  \frac{d}{dt} \sum_{i} {\tilde a}^i(t) &=& (S{\tilde H} 
S^{-1})^i_j {\tilde a}^j(t)\\
{}&=& (S^2 H S^{-2})^i_j  {\tilde \a}^j(t)
\eea
We now consult Sec. 4 of \cite{sv} and learn that $S^2 H S^{-2} =
H^\dagger$. Thus the string basis components evolve via
\be \label{comp3}
i  \frac{d}{dt} \sum_{i} {\tilde a}^i(t) = (H^\dagger)^i_j {\tilde a}^j(t),
\ee
a curious and useful fact! 

To apply this fact we simply go back to the expressions for the
decay amplitude in (\ref{tricon} ~-~ \ref{eshif}) and observe
that these results remain unchanged if we replace every matrix
element $V^i_j$ of the potential by that of the adjoint
$(V^\dagger)^i_j=V^j_i$. Thus the results obtained in Section 3 in
the gauge theory basis exactly describe the decay amplitude in
string
basis!

\section{Computation of the decay rate} 

In Section 3 we derived a formal expression for the decay rate that we
reproduce here,
\be\label{decay}
\G_n = 2\pi\int \ud\a_3 U^n_{\a_3} U_n^{\a_3}
\d(E_{\a_3}-E_n).
\ee
We explain the evaluation of this expression in some detail. Let us begin with 
the computation of the effective composite interaction matrices $U^n_{\a_3}$ 
and $U^{\a_3}_n$. 
Triple-trace state is labelled as $\a_3=\{m,s_0,s_1\}$. 
For convenience, let us also define $s_2=1-s_0-s_1$.
Because of the $\d$-function in (\ref{decay}), we need this
matrix element only at degeneracy, $E_{\a_3}=E_n$, where the
computation simplifies and has already appeared in \cite{umut2}
and \cite{golm1}.
Using the perturbation coefficients (\ref{Vcoef}) in (\ref{mel2}) one finds,
\bea\label{U1}
U_n^{ms_0s_1}&=& \frac{\l'g_2^2 n}{2\pi^4J}\sqrt{\frac{s_1s_2}{s_0}}
\int_0^1\ud s_0' \frac{1}{s_0'}
(\d_{s_0',s_0+s_1}+ \d_{s_0',s_0+s_2})\sin^2(\pi n s_0')\nonumber\\
{}&&\qquad\qquad\qquad\quad\sum_{p=-\infty}^{\infty}
\frac{p}{s_0'}\frac{\sin^2(\pi p s_0/s_0')}{(n^2-(p/s_0')^2)(n-p/s_0')^2}
\nonumber.
\eea
The evaluation of the sum in the second line is explained in
Appendix C of \cite{umut2}. 
Finally performing the integral over $s_0'$ one arrives at the result,
\be\label{U13}
U_n^{\a_3}=
-\frac{\l'g_2^2}{4\pi^2J}\sqrt{s_1s_2 s_0}\sin^2(\pi ns_1).
\ee
A similar computation using (\ref{Vcoef2}) in (\ref{mel1}) gives the
result\footnote{The factor of 2 arises from the $(k+1)!$ and $k!$
  prefactors in (\ref{Vcoef}) and (\ref{Vcoef2}).}
$U_{\a_3}^n = 2 U_n^{\a_3}$ when $E_{\a_3}=E_n$. 
Inserting (\ref{U13}) into (\ref{decay}) one obtains, 
\be\label{decay2}
\G_n=
\frac{\l'g_2^4}{4\pi^3 J^2}\sum_{m=-\infty}^{\infty}J\int_0^1\ud s_0\,\,
J\int_0^{1-s_0}\ud s_1\, s_1\,s_2\, s_0\sin^4(\pi n s_1)\d(n^2
-(\frac{m}{s_0})^2).
\ee
We evaluate this expression for $n>0$ for simplicity but the final
expression will be valid also for $n<0$.  
The integral over $s_1$ can be done analytically with the result,
\be\label{decay3}
\G_n= \frac{\l'g_2^4}{128 n^2\pi^5}\sum_{m=1}^\infty\int_0^1\ud s_0\,\d(n^2
-(\frac{m}{s_0})^2)s_0(1-s_0)(15+4\pi^2n^2(1-s_0)^2).
\ee
Let us denote the solutions to the degeneracy condition,
$$n^2=\frac{m^2}{s_0^2},$$
by $\{m^*,\, s^*_0\}$. One then performs the integral over $s_0$ using the 
$\d$-function,
$$\d(n^2-(\frac m{s_0})^2)=\frac{m}{2n^3}\d(s_0-\frac{m}{n}).$$
This gives,
\be\label{decay4}
\G_n= \frac{\l'g_2^4}{256\pi^5 n^7}\sum_{m^*=1}^{n-1} {m^*}^2 (n-m^*)
(15+4\pi^2(n-m^*)^2).
\ee
Here range of the sum is set by positive solutions to the
degeneracy condition above. For example, when $n=3$ there are two
solutions $\{m^*,s_0^*\}=\{1,1/3\}$ and $\{2,2/3\}$. The sum in
(\ref{decay4}) is easily done for general $n$ and one arrives 
at the final result, 
\be\label{res}
\G_n= \frac{\l'g_2^4}{3840\pi^3 n^5}(n^2-1)(n^2+1+\frac{75}{4\pi^2}). 
\ee 
We observe that decay width vanishes for $n=\pm 1$ and it 
shrinks as the excitation
mode $n$ increases.   

\section{Discussion}

The viewpoint we have taken is somewhat simplified in that we
have incorporated the degeneracy of single- and triple-trace
operators $\co^J_n$ and $\co^{J_0,J_1,J_2}_m$ when $m=\pm n J_0/J$,
but we have ignored further degeneracies, such as that of
$\co^{J_0,J_1,J_2}_m$ with the five-trace
$\co^{J_0',J_1,J_2,J_3,J_4}_p$ when $p = \pm mJ'_0/J_0$. Indeed
the state $\co^{J_0,J_1,J_2}_m$ is stable only when $m= \pm 1$,
so our calculation gives the true order $g_2^2$ 
amplitude for decay of $\co^J_n$ to $\co^{J_0=J/|n|,J_1,J_2}_{m= \pm
1}$. The rate is given by the $m^*=1$ term in the sum
(\ref{decay4}).
For $|n|>2$ and $|m|>2$, one must
envisage a sequential decay process, \eg  $$\co^J_{n=3} ~\to~
\co^{J_0,J_1,J_2}_{m=2}~\to~\co^{J_0',J_1,J_2,J_3,J_4}_{p=1}$$
with $ J_0/J = 2/3$ and $J'_0/J_0 = 1/2$. For general n, there
can be a cascade up to $(2|n|-1)-$trace states. We have not studied
transition amplitudes such as  $3~\to~ 5$ or sequential decays
explicitly, but we expect that the amplitude of the
$1~\to~3~\to~5$ process is of order $g_2^4$. Thus
we believe that the order $g_2^2$ decay amplitude we have calculated
is meaningful for general $n,m$ and that it can be readily compared
with string field theory calculations.

Readers should note that most statements made in this paper about
vanishing amplitudes hold for the lowest order contributions
only. One expects non-vanishing higher order corrections. For
example, the decay amplitude for $1~\to~2$ of order $g_2^3$ is
expected to be non-vanishing at degeneracy, and there should be a
non-vanishing $1~\to~4$ amplitude of the same order. On the other
hand the elementary order $g_2$ amplitude $V_{\a_3}^{\a_4}$ in
(\ref{Vcoef}) vanishes at degeneracy because it is disconnected,
{\it viz.} the delta functions in (\ref{delta}). 

As we stated in the Introduction, the computation of anomalous
dimensions of BMN operators requires degenerate perturbation
theory with possible modification of the order $g_2^2$ result
(\ref{eig1}). We now briefly describe our
preliminary study of this question in which we work at large
finite $J$. In general there are order
$g_2^2$ transition amplitudes which do not vanish at degeneracy between
a single trace operator $\co^J_n$ of momentum $n$ and triple-trace 
operators with several values of $m$. The triple-traces in turn
mix with 5-trace operators, 5-trace with 7-trace, etc., with
termination only at the maximal $J$-trace level. Note that each
``band'' of $k$-trace operators involves a finite
fraction of $J$ distinct operators. Nevertheless one can derive the 
precise statement that to order $\l' g_2^2$ the anomalous dimensions
of operators in this large set are the eigenvalues of
$H_0 + \l'g_2^2U$ where $U$ is the effective composite interaction
whose non-vanishing matrix elements are,
\be \label{compi}
U^{\a'_{k \pm 2}}_{\a_k} = \int \ud\a_{k \pm 1} \frac{V^{\a_{k \pm
2}}_{\a_k \pm 1} V^{\a_{k \pm 1}}_{\a_k}}{E_{n\a_{k \pm 1}}},
\qquad
U^{\a'_k}_{\a_k} = \sum_{j=\pm 1}\int \ud\a_{k+j} \frac{V^{\a'_k}_{\a_{k+j}} 
V^{\a_{k+j}}_{\a_k}}{E_{n\a_{k+j}}}.
\ee
Note that single-triple matrix elements appeared in Sec. 3. For large
$J$ this is a very large but sparse matrix. 

We have studied a reasonably
accurate ``model'' of this matrix to determine the large $J$
limit of eigenvalues and eigenvectors. The results indicate that
there is a single eigenvector which is ``mostly single trace'' as
$g_2^2 \to 0$, and its eigenvalue is that of (\ref{eig1}) for
$k=0$. Similarly, there are some ``mostly k-trace'' eigenvectors
for which  (\ref{eig1}) is also correct. But there are also
some ``collective'' eigenvectors which are superpositions of many
multi-trace operators and whose eigenvalues do not agree with
(\ref{eig1}). Thus our model indicates that (\ref{eig1}) is
correct for most states in the system but not correct for all
multi-trace states. 

We close by noting that the gauge theory result (\ref{eig1}) for
``mostly k-trace'' operators has not yet been confirmed in light
cone string field theory. The complication of
degenenerate perturbation theory described above in the gauge
theory will be mirrored in string theory. Thus we expect that the
string theory computation can be organized to produce a composite
interaction matrix at degeneracy which should agree with
(\ref{compi}) after change to string basis.

\centerline{\bf Acknowledgements}

We are indebted to David Berenstein who suggested the basic
approach taken in this research and made useful comments on the
preliminary manuscript. We also thank Marc Spradlin and Anastasia
Volovich for reading the manuscript and acknowledge useful
discussions with Neil Constable, Alfred Goldhaber, and Lubos Motl.
DZF's research is supported by the National Science Foundation
Grant PHY-00-96515 while both authors enjoy the support provided by
the D.O.E. under cooperative research agreement \#DF-FC02-94ER40818.

\end{document}